\documentclass[12pt]{iopart}
\usepackage{amssymb,latexsym,mathrsfs}
\usepackage{amsfonts}
\usepackage{graphicx}
\usepackage{color}    
\usepackage{textcomp}
\usepackage{iopams}

\newcommand{\be}{\begin{equation}}
\newcommand{\ee}{\end{equation}}
\newcommand{\bearr}{\begin{array}}
\newcommand{\enarr}{\end{array}}

\def\bea{\begin{eqnarray}}
\def\eea{\end{eqnarray}}
\def\ba{\begin{array}}
\def\ea{\end{array}}

\begin{document}

\title{How motility affects Ising transitions}
\author{Chandraniva Guha Ray, Indranil Mukherjee, P. K. Mohanty}
\address{Department of Physical Sciences, Indian Institute of Science Education and Research Kolkata, Mohanpur - 741246, India.}
\ead{pkmohanty@iiserkol.ac.in}

\begin{abstract}
We study a lattice gas model   of   hard-core particles  on a  square lattice experiencing  nearest neighbour attraction $J$.  Each particle  has  an internal  orientation, independent of the  others, that point towards  one  of the four  nearest neighbour and    it can move  to   the neighbouring site  along that  direction with the   usual  Metropolis rate  if  the  target site  is vacant.   The internal orientation of the particle can also  change to  any of the other  three  with a  constant rate $\omega.$  The dynamics of the model    in $\omega\to \infty$  reduces to   that of the Lattice Gas  (LG)   which exhibits a phase separation transition  at  particle  density $\rho=\frac12$  and temperature $T=1,$ when the    strength of  attraction  $J$  crosses a threshold  value $\ln(1+ \sqrt{2}).$ This transition belongs to Ising universality class.  For  any finite $\omega>0,$   the  particles  can be considered as  attractive run-and-tumble particles (RTPs) in two dimensions  with  motility $\omega^{-1}.$   We find that  RTPs also exhibit a  phase separation transition, but the critical  interaction  required  is $J_c(\omega)$   which  increases  monotonically   with increased  motility $\omega^{-1}.$   It appears that the transition   belongs to Ising universality class.  Surprisingly,  in these models, motility impedes cluster formation process   necessitating   higher interaction  to stabilize microscopic clusters.  Moreover,  MIPS like phases are not   found   when $J=0.$
\end{abstract}

\maketitle

\section{Introduction}
Ising model \cite{Onsager_Ising}  stands as a cornerstone  of   equilibrium  statistical mechanics offering  valuable insights into our understanding of critical phenomena, symmetry breaking,  phase transitions, renormalization group theories  and universality hypothesis.  Its  particle-conserved  counterpart, formally known as  the  lattice gas (LG) model, provides a simplified representation of a fluid or gas system. In LG, particles are confined to lattice sites and allowed to move following  the Metropolis-rate  corresponding to  nearest-neighbor attractive interactions (similar to Ising model).  For any nonzero  attractive  interaction,  the model exhibits an equilibrium phase transition  from a  homogeneous  mixed phase  to a  phase-separated state as  temperature $T$ of the system  is  lowered  below  a critical  value $T_c;$  both the  critical  temperature  and  the  critical exponents  of the model are  exactly known.  Attempts have been made to  study   the fate  of  these models  under non-equilibrium conditions \cite{KLS1,KLS2,TwoTemp,NoneqIsing} with a generic aim   to understand non-equilibrium phase transitions. One simple  way is  to use an external driving force or energy gradient leading to a biased movement  of particles along the  direction  opposite to the gradient.  It is well known that the phase separation   transition  of   driven lattice gas (DLG)  \cite{KLS1,KLS2} also belongs to  the  Ising universality class  with  a dynamical exponent different from its  equilibrium counterpart.  

In this article we  ask what happens  when  particles   in   a lattice gas  model do persistent motion?  We assign  each  particle  an  internal  sense of direction  which  points to one of  the nearest  neighbour. Now a particle   can move   with Metropolis rate, but  only    to a neighbouring   site  along  its  internal orientation, when that  site  is vacant.  In addition,  they are  allowed  to  change their  internal direction  and reorient  along one of the  other three  directions  with  a constant rate  $\omega.$  Note that   the directional preference  generates persistent  run  of particles in a  specific direction   followed by a tumbling of direction.  Such particles, formally known as the run and tumble particles   (RTPs)   constitute an important class of particles  in active matter systems.

Active  particles   generally  consume energy from the environment to produce persistent  self-propelled motion \cite{Rev1, C7, C9, C10, Bechinger2016, B10, Tailleur_Rev2022}, commonly referred to as motility.   Motion of these particles  are   categorized into  two  broad  classes. Active Brownian  particles (ABPs) which run  in specific directions  but  their direction  changes  continuously following  noisy dynamics. In contrast,  run and tumble particles   execute   unidirectional runs  between two tumbling events  that  happen with a  constant rate.  Active collective motion is common in the animal world starting from flocking of birds \cite{B7}, schooling of fish \cite{B6}, swarming of insects \cite{B8} in macro-scale to cell migration \cite{B10}, and crowding of bacteria \cite{B12, C59} in micro-scale. Some bacteria and algae \cite{A5, A6} perform a specific kind of self-propelled motion -- a sequence of persistent runs along a specific direction followed by `tumbling' (change of orientation) \cite{C46, A, cates_motility-induced_2015}.

Usually, the interplay between random fluctuations and persistent propulsion in these systems result in nonequilibrium steady states that exhibit collective behavior at many different length scales. It is well  documented that   persistence driven by activity  leads  to  jamming   or phase co-existence; transition from  an isotropic homogeneous  phase to  a mixed phase of co-existing high and low-density   have been observed  with  increased  persistence \cite{C46, C47, C48, C49, C50, C51, C52, C53, C54}. Besides  numerical investigation \cite{Redner, Reentrant_I, Chate2020, Reentrant_II, Reentrant_III, Tailleur_Rev2022},  theoretical  
study of MIPS have thus far concentrated on hydrodynamic descriptions of the coarse-grained local density \cite{cates_motility-induced_2015, C9, C46}, 
agent based modeling  \cite{fily_athermal_2012, C9, bialke_microscopic_2013},  and   lattice models  \cite{thompson_lattice_2011, slowman_jamming_2016,mallmin_exact_2019, dandekar_hard_2020,NoMIPS_1D, Golestanian, Whitelam, 
Solon_Tailleur_2015, Soto-2016}. 

 \begin{figure}[h]
\centering 
\includegraphics[width=10.cm]{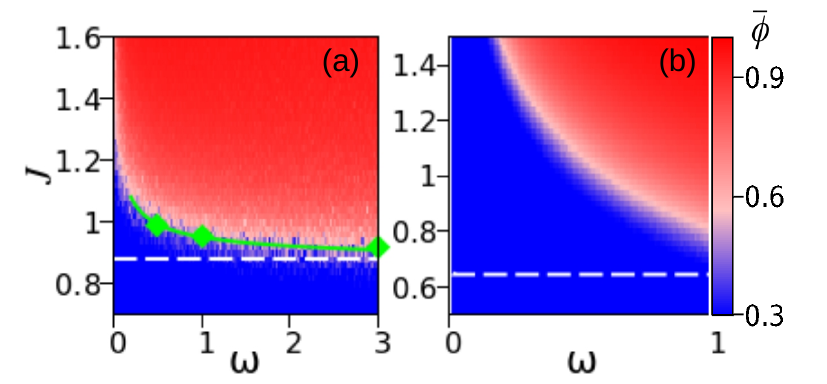}
\caption{Density plot of order parameter $\bar{\phi},$ averaged over more than $10^6$ samples, in $\omega$-$ J$ plane  with grid size $0.02$ in both axes. (a) Conserved lattice gas model ($64\times 32$ square lattice) of RTPs. (b) Infinitely driven lattice gas ($32\times32$ square lattice) model of RTPs. The solid line in (a) corresponds to the critical line $J_c(\omega),$ estimated from the best fit of the contour line that passes through the three accurately determined critical points (symbols), $J_c=0.993, 0.956, 0.919$ for $\omega = \frac12,1,3$ respectively.   Since the model is ill-defined  when $\omega=0,$ we could not decide  whether $J_c \to \infty$ or   it remains finite in $\omega\to0$ limit. 
It is evident from the color gradients that $\bar{\phi},$ a measure of order due to clustering, decreases in both models when motility $\omega^{-1}$ is increased. 
Dashed lines (a) $J^{LG}_c=0.881$, (b) $J^{DLG}_c=0.650$ represent the critical values of $J$ known \cite{Mx2-My2} for $\omega\to \infty.$ Clearly, MIPS transition does not occur in these models when $J=0.$}
\label{fig:heat_map}
\end{figure}

In this article, we focus on hardcore particles  on a two dimensional (2D) square lattice with  attractive inter-particle interaction. In absence  of  activity,  particles in this model move to  their neighbouring vacant sites with   the standard Metropolis rate. This model,    formally  known as   the  conserved  lattice gas (LG) model,  exhibits a  phase separation transition   
when interaction $J$ is  increased beyond  a threshold  - for  particle density $\rho=\frac12$  the  transition occurs at  $J_c= \ln (1+\sqrt2)T.$   We introduce activity  in the  model by associating  an internal  orientation  vector  that  can point along one of  the nearest neighbours - particles  are then  allowed to move  only along  this direction  with    Metropolis rate and they can  reorient   to another   direction  with a constant rate $\omega.$  Since   in free space,  the probability  that a particle  
runs  along  their internal direction  without being  reoriented (tumbled)  up to time $t$ is   $e^{-\omega t},$ one  may consider  the   persistent  length of these run and tumble particle  to be $\omega^{-1}.$   When $\omega \to \infty,$  particles  tumble infinitely many times before  making   a move - which is similar to  the  usual Metropolis   dynamics  where  particles move  to   neighbour  chosen  randomly and independently.  Thus  interacting  run-and tumble particle (IRTP) model  reduces  to  the usual LG model    in $\omega \to \infty$ limit  which exhibits a phase separation transition belonging to Ising universality class (IUC). 
Our  primary aim is to  see how  motility $\omega^{-1}$  affects  the   phase separation  transition.

Our findings reveal that, regardless of motility magnitude, phase separation only occurs when the attractive interaction among RTPs surpasses a finite threshold,  which is  larger than $\ln (1+\sqrt2);$ this transition remains  in  Ising universality class (IUC). Furthermore, we observe that stronger interaction  is required  to order  the system when  motility is larger (i.e., the critical interaction strength $J_c(\omega)$ increases monotonically with increasing motility). This trend,  that increased motility impedes cluster formation, is found to be  consistent across  other  models, exemplified by (a) interacting lattice gas RTPs, (b) infinitely driven lattice gas RTPs.  In all cases, the order parameter decreases with rising motility ($\omega^{-1}$), as evident from the  density plot of the order parameter  shown in Fig. \ref{fig:heat_map}.

\section{Interacting RTP Model}

Let us consider $N$ hardcore RTPs on a square lattice lattice ${\cal L}$ with periodic boundary conditions in both directions, where sites are labeled by ${\bf i}\equiv (x,y)$ with $x=1,2 \dots, L_x$ and $y=1,2 \dots, L_y.$ Each particle $k=1,2, \dots, N$ carries an internal orientation vector ${\bf s}_k \in \{ {\bdelta}_0, {\bdelta}_1,{\bdelta}_2, {\bdelta}_3 \}.$ 
${\bdelta}_l \equiv (\cos \frac{\pi l}2, \sin \frac{\pi l}2)$ 
with $l=0,1,2,3,$ are unit vectors pointing to the nearest neighbors of a site, on a square lattice. Respecting the excluded volume or hardcore nature of the RTPs, each site ${\bf i}$ is allowed to accommodate at most one particle, thus the occupancy $ n_{x,y}\equiv n_{\bf i} = 0,1$ and 
$\sum_{\bf i} n_{\bf i} = N.$ We introduce attractive inter-particle interaction among RTPs using an energy function 
\be
 E(\{n_{\bf i}\})=-2J\sum_{{\bf i} \in {\cal L}} \sum_{l=0}^3 n_{\bf i } n_{{\bf i} +{\bdelta}_l}.
 \label{eq:IsingE}
 \ee
\begin{figure}[h]
\vspace*{.1 cm}
\centering
\includegraphics[width=0.5\textwidth]{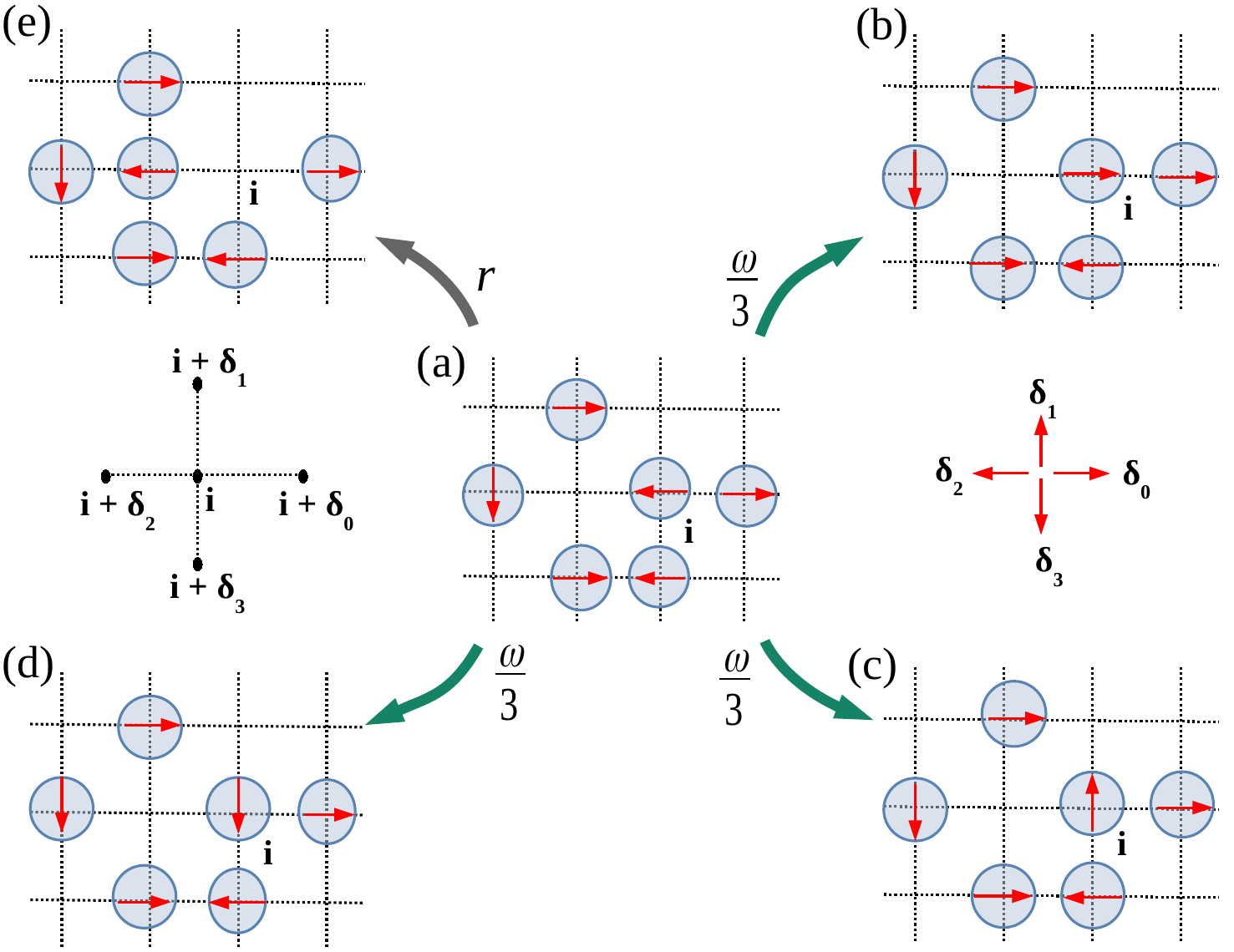}
\caption{Dynamics of IRTP model. (a) The particle at site ${\bf i}$ has the internal orientation ${\bdelta}_2=(-1,0).$ 
It may tumble with rate $\omega$ and reach
any of the configurations (b), (c), or (d), each one with rate $\omega/3.$ It may also move along ${\bdelta}_2$- direction with the Metropolis rate $r= {\rm Min}\{1, e^{-\Delta E}\}= e^{-2J}$ from the site ${\bf i}$ to site ${\bf i} + {\bdelta_2}$ and reach the configuration (e).}
\label{fig:dynamics}
\end{figure}
The dynamics of RTPs are as follows: the particle $k$  (at site ${\bf i}$) may tumble and change its orientation ${\bf s}_k$ to any one of the other three with rate $\omega$ or it may run to the neighboring site ${\bf i} + {\bf s}_k $ along the direction ${\bf s}_k$ with the usual Metropolis rate $r= {\rm Min}\{1, e^{-\Delta E}\}$
only if $n_ {{\bf i} + {\bf s}_k}=0.$ Here, $\Delta E$ is the energy difference of the target configuration with respect to the initial one, calculated using Eq. (\ref{eq:IsingE}).

The dynamics of the model are described schematically in Fig. \ref{fig:dynamics}. In this example, the RTP at site ${\bf i}$ has an orientation ${\bdelta}_2$. It 
can move to the neighbor ${\bf i}+{\bdelta}_2$ with rate $ r = e^{-2J},$ where $\Delta E = 2J$ from Eq. (\ref{eq:IsingE}). It can also tumble and reorient itself to one of the directions ${\bdelta}_0, {\bdelta}_1,$ or $ {\bdelta}_3$ as illustrated in Fig. \ref{fig:dynamics} (b), (c) and (d).

Note that  the restriction that   RTPs move  only along their  internal orientation   leads to  violation of   the  detailed balance condition and the steady-state of the system   different from   the  Boltzmann
distribution  {\it w.r.t}  the  energy function given by 
Eq. (\ref{eq:IsingE}).  However, the choice of run rate  that  resembles  the   Metropolis  rate {\it w.r.t} an energy function given by 
Eq. (\ref{eq:IsingE})  has an advantage. In the  $\omega \to \infty$ limit, RTPs tumble infinitely many times before attempting a run which is equivalent to an ordinary particle choosing one of the four directions randomly. Thus, in this limit, the RTP dynamics becomes the usual Kawasaki dynamics of LG at temperature $T=1,$ which exhibits a phase separation transition at $J^{\rm LG}_c=\ln(1+ \sqrt2)$ \cite{Book_Dickman}. Thus,
\be \lim_{\omega\to\infty} J_c(\omega)= J^{\rm LG}_c=\ln(1+ \sqrt2)\simeq 0.881
\label{eq:TcLG} \ee

A few comments are in order. First, RTPs on a lattice move only one lattice unit at a time with a rate $r.$ The mean distance they run in $\Delta t $ time is then $r\Delta t$ and the mean speed is $v=r.$ Note that $r$ depends on  energy difference  between transiting configurations and  thus the mean  speed $v$ in the steady state  depends  on the  density of the system.  Secondly, the energy $E(.)$ does not depend on $\{{\bf s}_k\}.$ Thus a constant tumble rate $\omega$ leads the system to a steady state where all four orientations are equally likely,  leading to no global orientational order. 
 
\subsection{ The order parameter}
\begin{figure}[h]
\vspace*{.1 cm}
\centering
\includegraphics[width=0.7\textwidth]{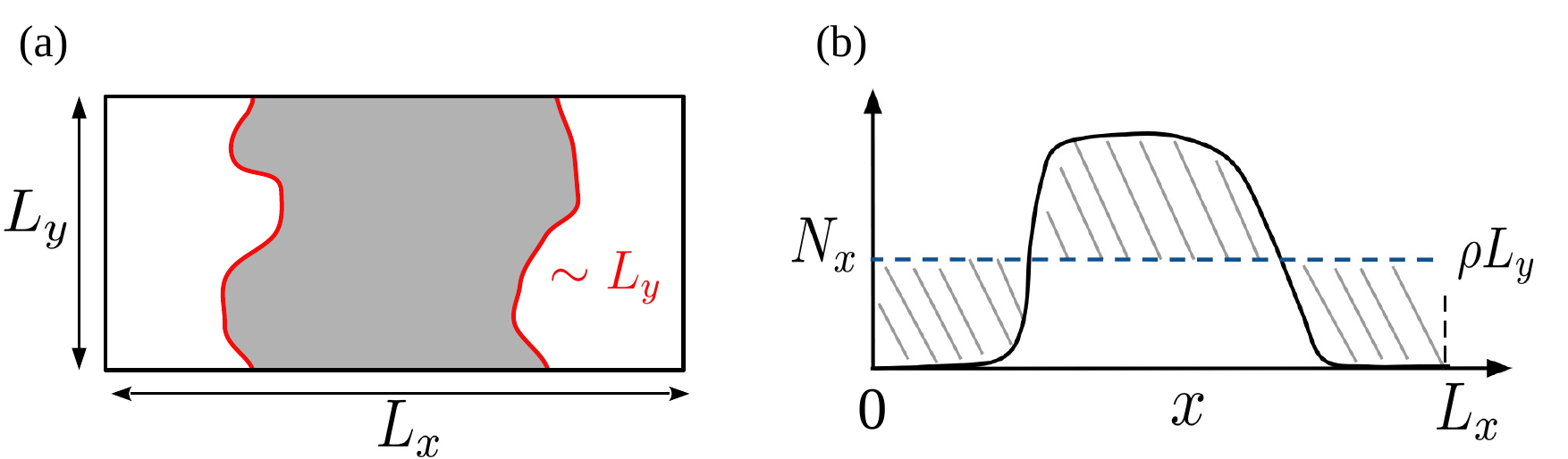}
\caption{ (a) Schematic configuration of a 
phase-separated state on a rectangular lattice $(L_x=2L_y).$ The coexistence line depicted in red has length $\sim L_y.$ 
(b) The order parameter of the system measures how different is $N_x$ from its mean $\rho L_y$ in an absolute sense (the shaded area). Here $N_x$ counts the total number of particles at lattice sites ${\bf i}\equiv (x,y)$ with the same $x$-coordinate. }
\label{fig:orderparameter}
\end{figure}

To proceed further, we need an order parameter that suitably describes the phase separation transition. For systems 
where the coexistence line 
(the line separating the high and low-density regions)  aligns in a preferred direction (say $y$), the order parameter  can be defined a \cite{Mx2-My2, Albano, Urna} as 
\be 
\centering
\phi = \frac{2}{L_x L_y} \sum_{x=1}^{L_x} \left| N_{x} - \rho L_y\right|;~ N_{x}= \sum_{y=1}^{L_y} n_{x,y}. 
\label{eq:OP}
\ee
which calculates how different is $N_x$ from its mean $\rho L_y$ in an absolute sense (the shaded area in Fig. \ref{fig:orderparameter}), where $N_x$ counts the total number of particles at lattice sites ${\bf i}\equiv (x,y)$ with the same $x$-coordinate. Also, for a  disordered (homogeneous)  system  $\phi \to 0$  in the thermodynamic limit. 

First, we verify that the steady state average  $\langle \phi\rangle$ (denoted as  $\bar \phi$)  correctly reproduces the known critical behavior of  non-motile  particles.
In the absence of motility, the IRTP model reduces to the well-known conserved LG  model which exhibits a phase separation transition at $J^{LG}_c=\ln(1+ \sqrt2)\simeq 0.881$ when temperature $T=1$ and particle density $\rho=\frac12$. The critical exponents of LG are known to be $\beta=\frac18, \gamma=\frac74,\nu=1$, belonging to the Ising universality class (IUC) in 2D. To verify  this  critical  behaviour we perform Monte-Carlo simulations of the LG  model  on a rectangular system ($L_x=2L, L_y= L$) considering  $\phi$  in Eq. (\ref{eq:OP}) as the  order-parameter. At  density  $\rho=\frac12$,  we  obtain its steady state average  $\bar \phi,$ and the susceptibility $\chi =\langle \phi^2\rangle - \langle \phi\rangle^2,$ as a function $J.$  To locate the critical value $J_c(\omega)$, we use the Binder cumulant ratio 
\begin{equation}
Q_L=\frac{\langle \phi^2\rangle ^2}{\langle \phi^4\rangle},
\label{eq:BC}
\end{equation} 
\begin{figure}[h]
\centering
\includegraphics[height=4.1cm]{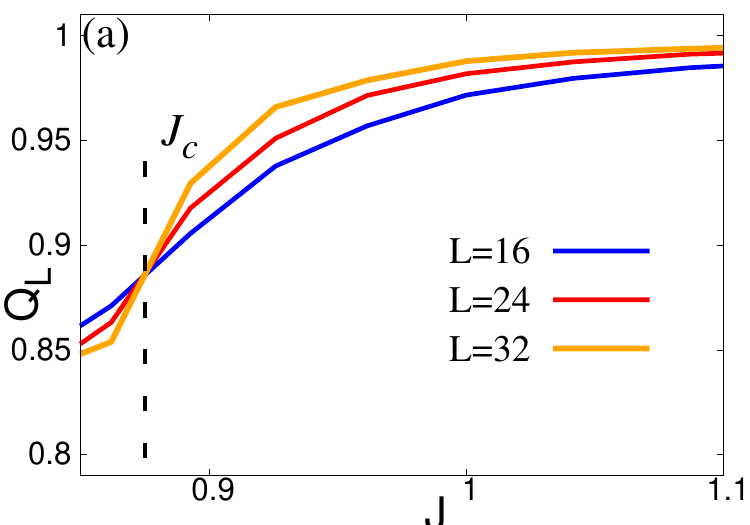}
\includegraphics[height=4.1cm]{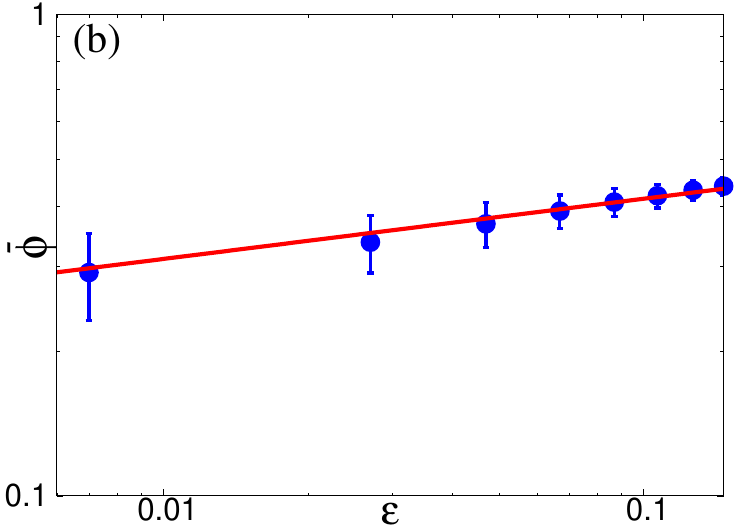}
\includegraphics[height=4.1cm]{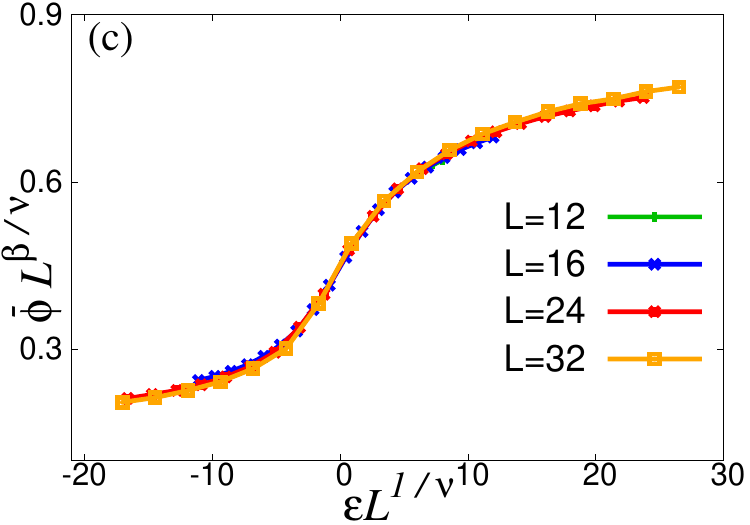}
\includegraphics[height=4.1cm]{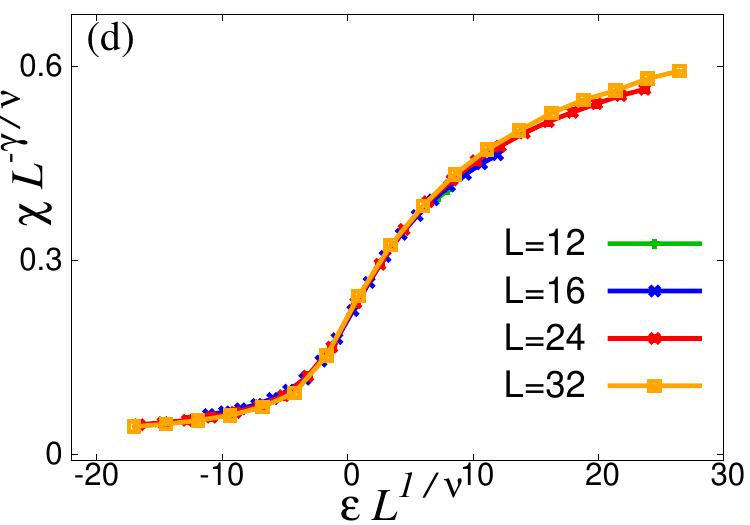}
\caption{ Critical behavior of LG: (a) Variation of the Binder cumulant $Q_L$ with interaction of strength $J$ for $L=16,24,32$; the intersection point is the critical temperature $J_c=0.881 \pm0.005$, (b) Log scale plot of $\bar{\phi}$ as a function of $\varepsilon = (J_c^{-1}-J^{-1})$ for $L=32.$ A solid line with slope $\beta=\frac{1}{8}$ is drawn for comparison.
Plot of (c) $\bar{\phi} L^{\beta/\nu}$ and (d)$\chi L^{-\gamma/\nu}$ as a function of $\varepsilon L^{1/\nu}$ for $L=16, 24,32$ 
 shows a good data collapse when Ising critical exponents $\frac{\beta}{\nu}=\frac18$ and $\frac{\gamma}{\nu}=\frac74$ are used. In each case, statistical averaging is done for more than $10^8$ samples.}
 \label{fig:ising_graphs}
\end{figure}
which is  known to be  independent of the system size  $L$ at the critical point \cite{BinderCum, Binder1, Binder2,6l}. $Q_L$ vs. $J$ curves obtained from Monte Carlo simulation for $L=16,24,32$ are shown in \ref{fig:ising_graphs}(a); they intersect 
at $J_c=0.881,$ which is is in good agreement with the known value $J_c^{\rm LG}= \ln (1+ \sqrt 2)$ for $T=1.$ 

 Further, the critical exponents  $\frac{\beta}{\nu}$, $\frac{\gamma}{\nu}$ and $\frac1\nu$  are obtained from finite-size scaling
\cite{FSS1, BinderBook, FSS2}, 
\begin{eqnarray}\label{eq:scaling_binder}
 \bar{\phi} = L^{-\beta/\nu} f_{\phi}(\varepsilon L^{1/\nu}); ~~ 
 \chi= L^{\gamma/\nu}f_{\chi}(\varepsilon L^{1/\nu}).
\end{eqnarray}
where $\varepsilon = J_c^{-1}-J^{-1}.$
Figures \ref{fig:ising_graphs}(c) and (d)
respectively show plot of $\bar{\phi} L^{\beta/\nu}$ and $\chi L^{-\gamma/\nu}$
as a function of $\varepsilon L^{1/\nu}$ for $L=16, 24,32.$ The Ising exponents $\beta= \frac18, \nu =1, \gamma = \frac{7}{4}$ provide a good data collapse indicating that the phase separation transition indeed belongs to IUC. Note that the dynamical exponent of the conserved Ising model (in 2D) is $z= \frac{15}4;$ this slows down the relaxation process drastically, particularly near the critical point. This restricts us from simulating larger systems. 

\subsection{Phase transition in IRTP model}

\begin{figure}[t]
\centering
\includegraphics[height=4.1cm]{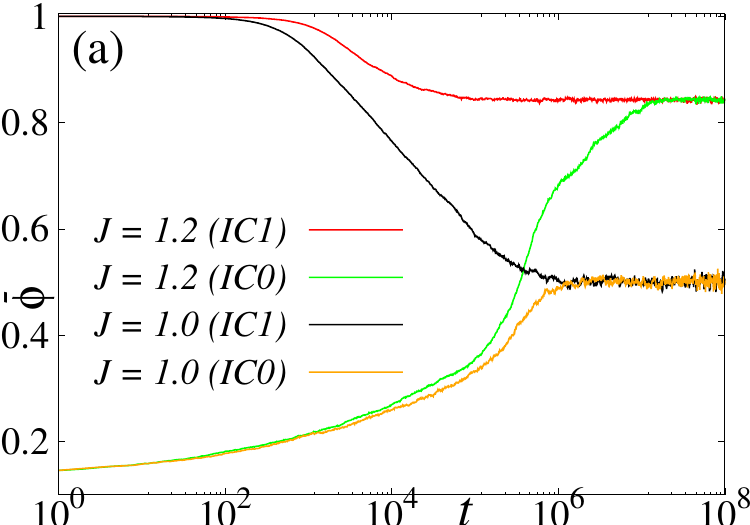}\hspace{0.3cm} 
\includegraphics[height=4.1cm]{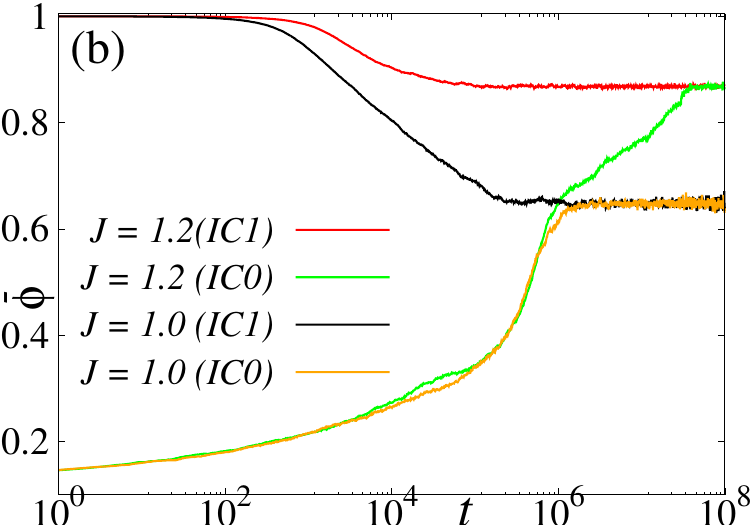}
\caption{ $\bar\phi(t)$  as a function of $t$ for (a) $\omega=0.5$ and (b) $\omega=1.0.$  For each $\omega,$
we consider  $J=1.0, 1.2$   which are  larger than $J_c,$ and two initial conditions,  IC1: all particles tightly packed  in  a $L\times L$ square  and   IC0:  particles placed  randomly on a lattice .  Each particle  is   assigned an  orientation chosen randomly from the set of unit vectors $\{ {\bf \delta_1, \delta_2, \delta_3, \delta_4 }\}.$  System size considered here is $L=32$ and  statistical averaging is done  over $300$ or more runs.}
\label{fig:diff_init_RTP_time}
\end{figure}

For interacting  RTPs, we  repeat the  Monte-Carlo simulations of on a rectangular system ($L_x=2L, L_y= L$) at  $\rho=\frac12,$  keeping in mind that  particles move   only along their internal orientation. In addition   particles can change their internal orientation   with rate $\omega$  and   reorient  to one of the other three directions.    First we check that  the  order parameter $\phi$ evolves to attain  a unique stationary  value  independent of the initial condition. Figure \ref{fig:diff_init_RTP_time} (a) and (b) represent the time-evolution of the  order parameter  $\bar \phi(t)$ for  different $\omega = 0.5, 1$ and
$J=1.0,1.2$  which are  in the phase separated state. 
As expected,   for a given $J,\omega,$  $\bar\phi(t)$  approach a constant  independent of the initial condition  as $t\to \infty.$ The relaxation time is larger   when  $J$ is closer to its critical value. Surprisingly, in all cases,  a fully ordered configuration (where particles are packed  in  a  $L\times L$  square)  relaxes much faster to  the steady-state  compared to  the random initial condition. 

\begin{figure}[h]
\centering
\includegraphics[height=4.1cm]{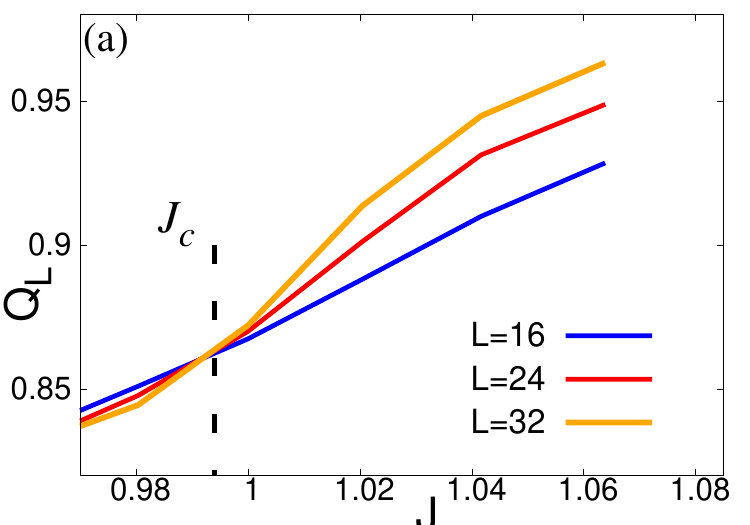}
\includegraphics[height=4.1cm]{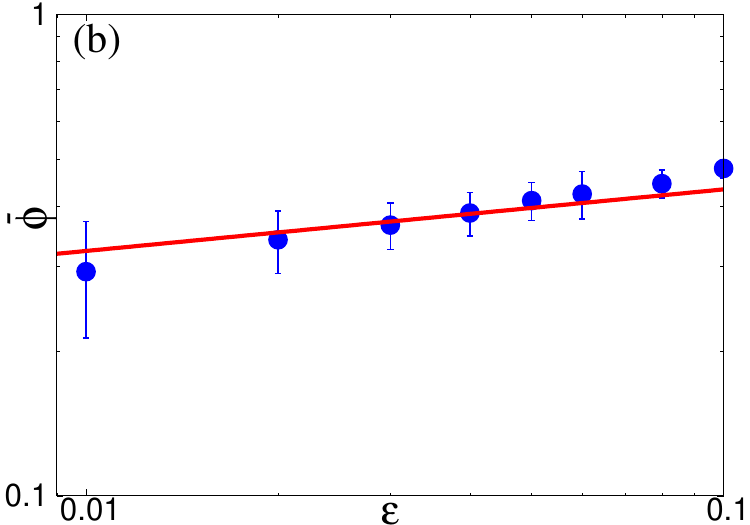}
\includegraphics[height=4.1cm]{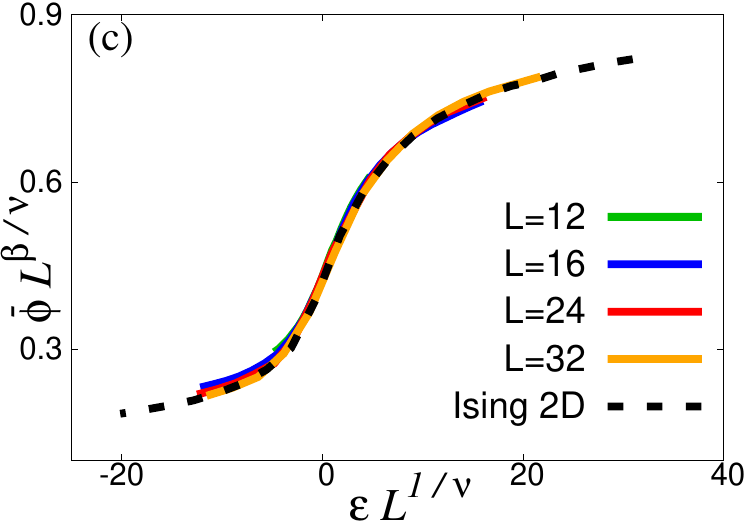}
\includegraphics[height=4.1cm]{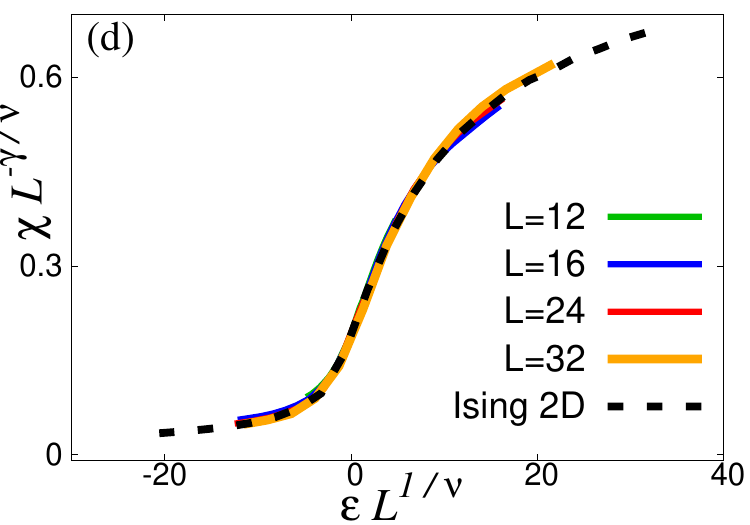}
\caption{IRTP with $\omega=0.5$: (a) Intersection point of Binder cumulants $Q_L$ for $L=16,24,32$ determines $J_c=0.993 \pm0.005.$ (b) Plot of $\bar{\phi}$ vs. $\varepsilon = (J_c^{-1}-J^{-1})$ in log-scale for $L=64$ (symbols) along with a line of slope $\beta= \frac18.$ (c) $\bar{\phi} L^{\beta/\nu}$ and (d) $\chi L^{-\gamma/\nu}$ as a function of $\varepsilon L^{1/\nu}$ for $L=12,16, 24,32$ exhibit scaling collapse for Ising critical exponent $\frac{\beta}{\nu}=\frac18$ and $\frac{\gamma}{\nu}=\frac74$. 
Dashed lines: respective scaling functions of Ising universality class.
Statistical averaging is done for more than $10^8$ samples.}
\label{fig:omega0.5_graphs}
\end{figure}

New we compute  the steady state average of order parameter $\bar \phi$ and the susceptibility $\chi =\langle \phi^2\rangle - \langle \phi\rangle^2$ 
are obtained as a function $J$ keeping $\omega$ fixed. 
From  the  Binder cumulant ratio $Q_L$ we locate the critical value $J_c$  and  then repeat the procedure for a different $\omega.$ 
For $\omega=0.5$   we estimate that $J_c= 0.993$ (Fig. \ref{fig:omega0.5_graphs}(a)). Figure \ref{fig:omega0.5_graphs}(b) presents a plot of $\bar \phi$ as a function $\varepsilon = J_c^{-1}-J^{-1}$ in log scale. Clearly, $\bar \phi \sim \varepsilon^\beta$ with Ising exponent $\beta=\frac18.$  Other critical exponents are obtained from  finite-size scaling as described in Eq.  (\ref{eq:scaling_binder}). In Figs. \ref{fig:omega0.5_graphs}(c)-(d) we plot 
$\bar{\phi} L^{\beta/\nu}$ and $\chi L^{-\gamma/\nu}$ 
as a function of $\varepsilon L^{1/\nu}$ for different $L.$ We observe a good data collapse using Ising exponents $\beta= \frac18, \gamma=\frac74, \nu=1.$ The functions $f_{\phi}(.)$ and $f_{\chi}(.)$ are also in excellent agreement with the universal scaling functions (dashed lines) of IUC with conserved dynamics \cite{Kawasaki}.

\begin{figure}[h]
\centering
\includegraphics[height=4.1cm]{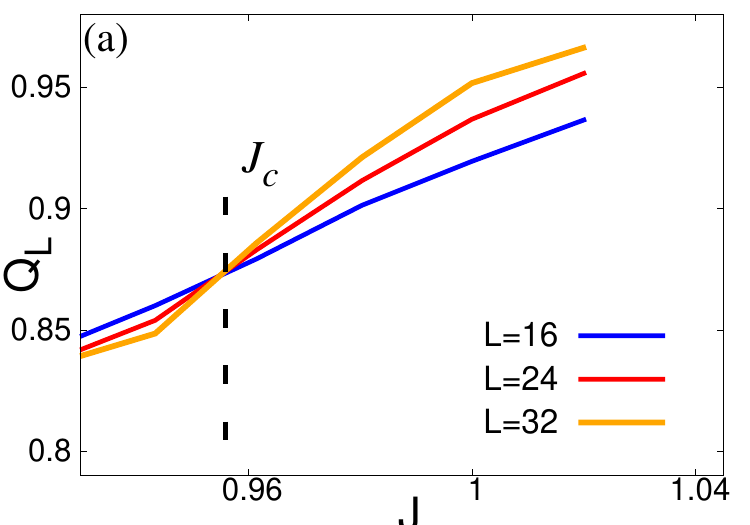}
\includegraphics[height=4.1cm]{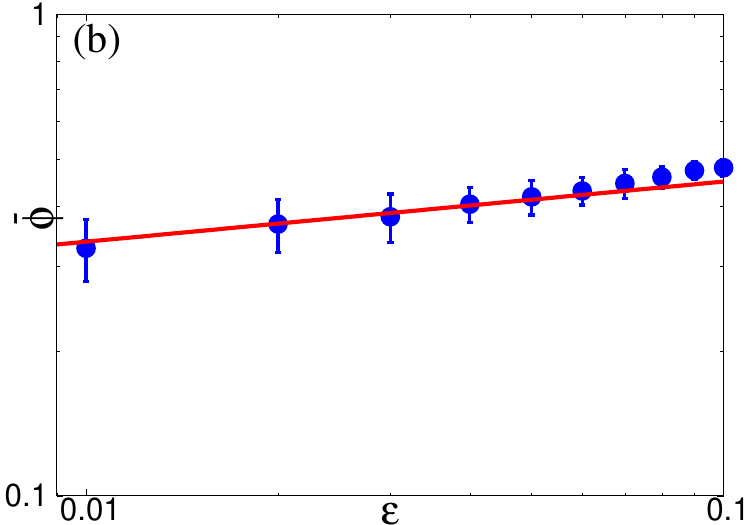}
\includegraphics[height=4.1cm]{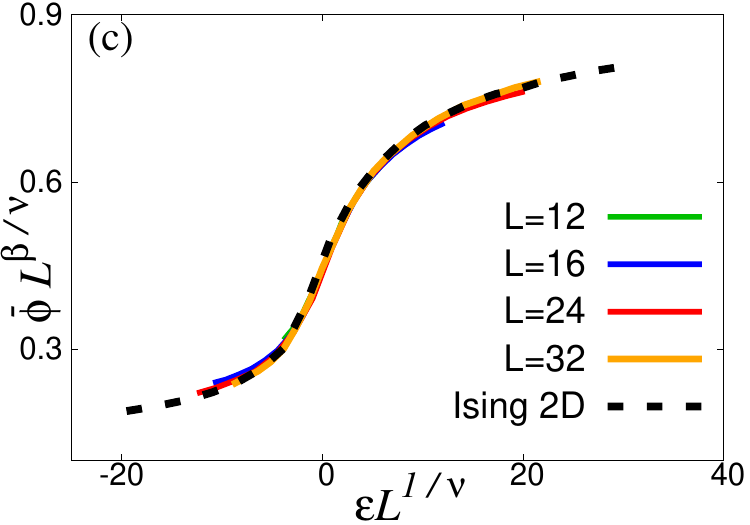}
\includegraphics[height=4.1cm]{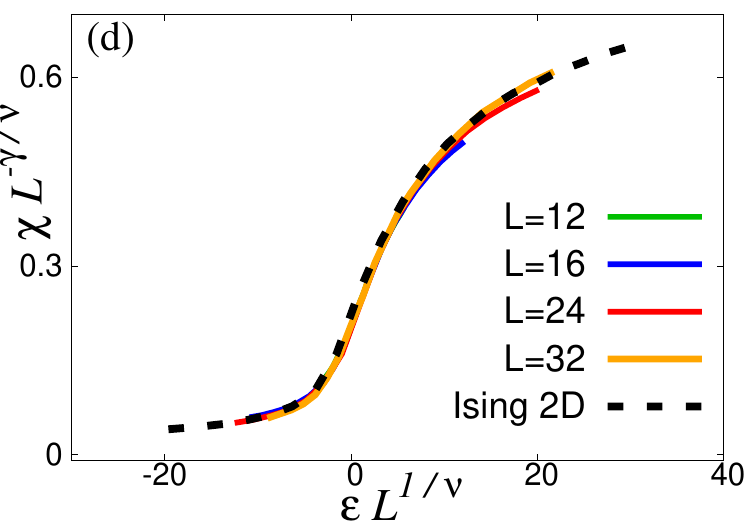}
\caption{ IRTP with $\omega=1.0$: 
The figures are identical to Fig. \ref{fig:omega0.5_graphs} except 
$J_c=0.956 \pm0.005.$}
\label{fig:omega1.0_graphs}
\end{figure}

\begin{figure}[t]
\centering
\includegraphics[height=4.1cm]{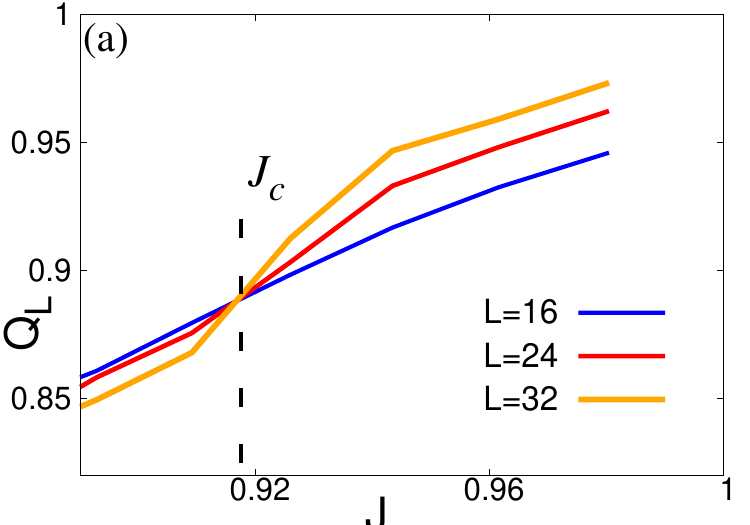}
\includegraphics[height=4.1cm]{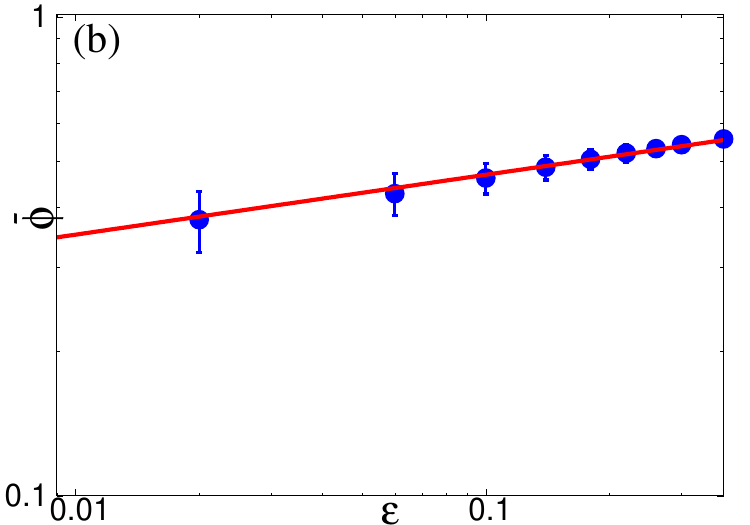}
\includegraphics[height=4.1cm]{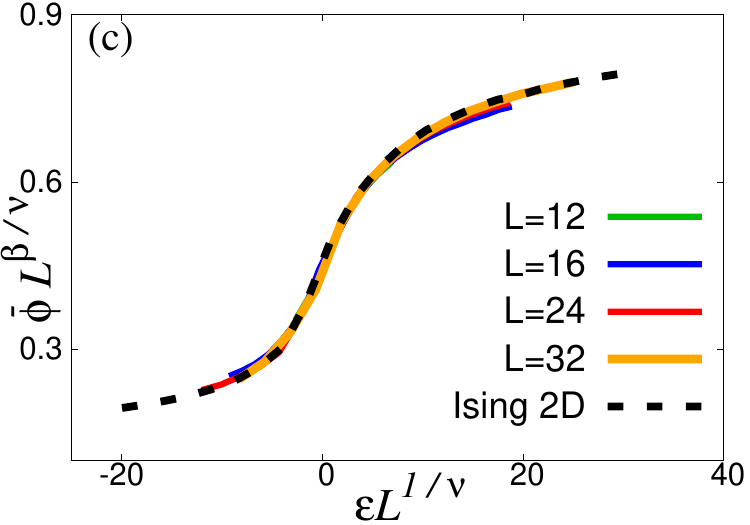}
\includegraphics[height=4.1cm]{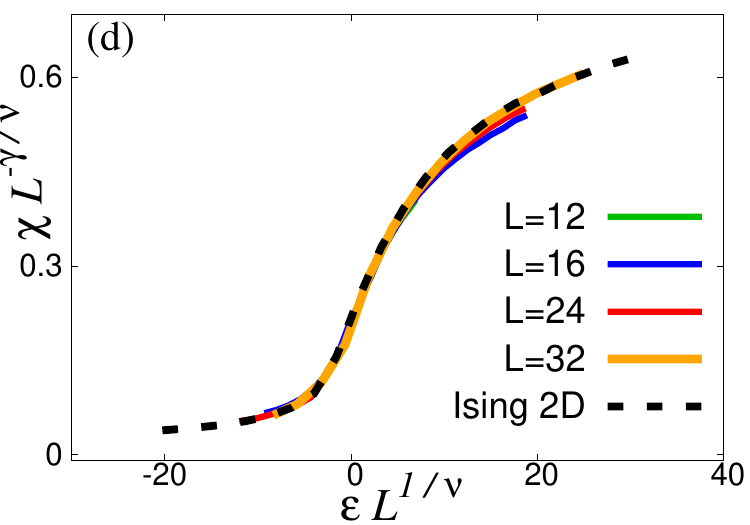}
\caption{IRTP with $\omega=3.0$: The figures are identical to Fig.  \ref{fig:omega0.5_graphs} except $J_c=0.919 \pm0.005.$
 }
\label{fig:omega3.0_graphs}
\end{figure}

We have also investigated the critical behavior at $\omega=1.0, 3.0$  in a similar way, respectively  in  Fig.  \ref{fig:omega1.0_graphs} and Fig. \ref{fig:omega3.0_graphs}.  The critical exponents and the scaling functions near the respective critical points $J_c=0.956, 0.919,$ are found to be consistent with IUC. We thus conclude that the phase separation transition produced by IRTPs belongs to IUC. This result is consistent with the critical behavior of other RTP models studied earlier \cite{MIPS-Ising2D1, MIPS-Ising2D2, MIPS-Ising2D3}.

Our concern is the observation $J_c(0.5)> J_c(1)> J_c(3)> J_c(\infty),$ which indicates that $J_c(\omega)$ might be a monotonically decreasing function of $\omega.$ To verify this, we need more accurate estimate of $J_c(\omega)$ for small $\omega$ values. This is computationally expensive, as the persistence length $\omega^{-1}$ and thus the relaxation time of the system diverges as $\omega\to 0.$ Instead, we calculate the $\bar \phi(\omega, J)$ from simulations and make a density plot in the $\omega$-$J$ plane, shown in Fig \ref{fig:heat_map}(a). It is evident from the color gradient that $\bar \phi$ decreases with the increase of motility $\omega^{-1}$. 
The curved line in Fig \ref{fig:heat_map}(a) provides a rough estimate of $J_c;$ it passes through the accurately estimated critical points obtained for $\omega=0.5,1,3.$ A dashed line 
$J^{LG}_c=0.881$ depicts the known critical value for $\omega\to \infty$ and emphasizes the fact that 
 $J_c(\omega)> J^{LG}_c ~ \forall ~\omega>0.$ 
 If motility helps phase separation one would need less attractive interaction to order a system when motility is large; we find the opposite.
Irrespective of the value of motility the system, in the  absence of positional diffusion, particles cannot phase separate unless there is finite attractive interaction $J> \ln (1+\sqrt2).$ This transition is rather induced by the interaction, not by motility.

\subsection{Other models}

{\it   Driven RTPs  in presence of attraction:}  Let us apply a bias to the interacting RTPs, say in the $+$ve $y$ direction, so that  their movement in ${\bdelta}_1$ direction is enhanced in comparison to ${\bdelta}_3.$ We study the infinite bias case where the run rate is set $r=1,0$ respectively for attempts in ${\bdelta}_1,$ ${\bdelta}_3$ directions. Run rates along ${\bdelta}_{0}$, ${\bdelta}_2$ are unaltered, $r= {\rm Min}\{1, e^{-\Delta E}\}.$ 
 In the absence of motility, i.e., when $\omega \to \infty,$ the dynamics of the model reduce to 
 that of infinitely driven lattice gas (DLG) 
 \cite{KLS1, KLS2, Book_Dickman, Mx2-My2, Albano, Urna} which undergoes a phase separation transition under nonequilibrium conditions at $J^{DLG}_c= J^{LG}_c/ \alpha$ with $\alpha= 1.355$ \cite{Mx2-My2}. For generic $\omega <\infty$, we obtain from Monte Carlo simulations the order parameter $\bar\phi,$ following Eq. (\ref{eq:OP}) for tumble rate $ \omega \in (0,1),$ and $ J \in (0.5, 1.5).$ A density plot in Fig. \ref{fig:heat_map}(b) shows that $\bar \phi$ decreases as motility is increased. We find that phase separation cannot occur in this system, no matter how large the motility is, unless the attractive interaction exceeds $J^{DLG}_c \simeq 0.650.$ 

\section{Conclusion and discussions}

\begin{figure}[h]
\centering
\includegraphics[scale=0.4]{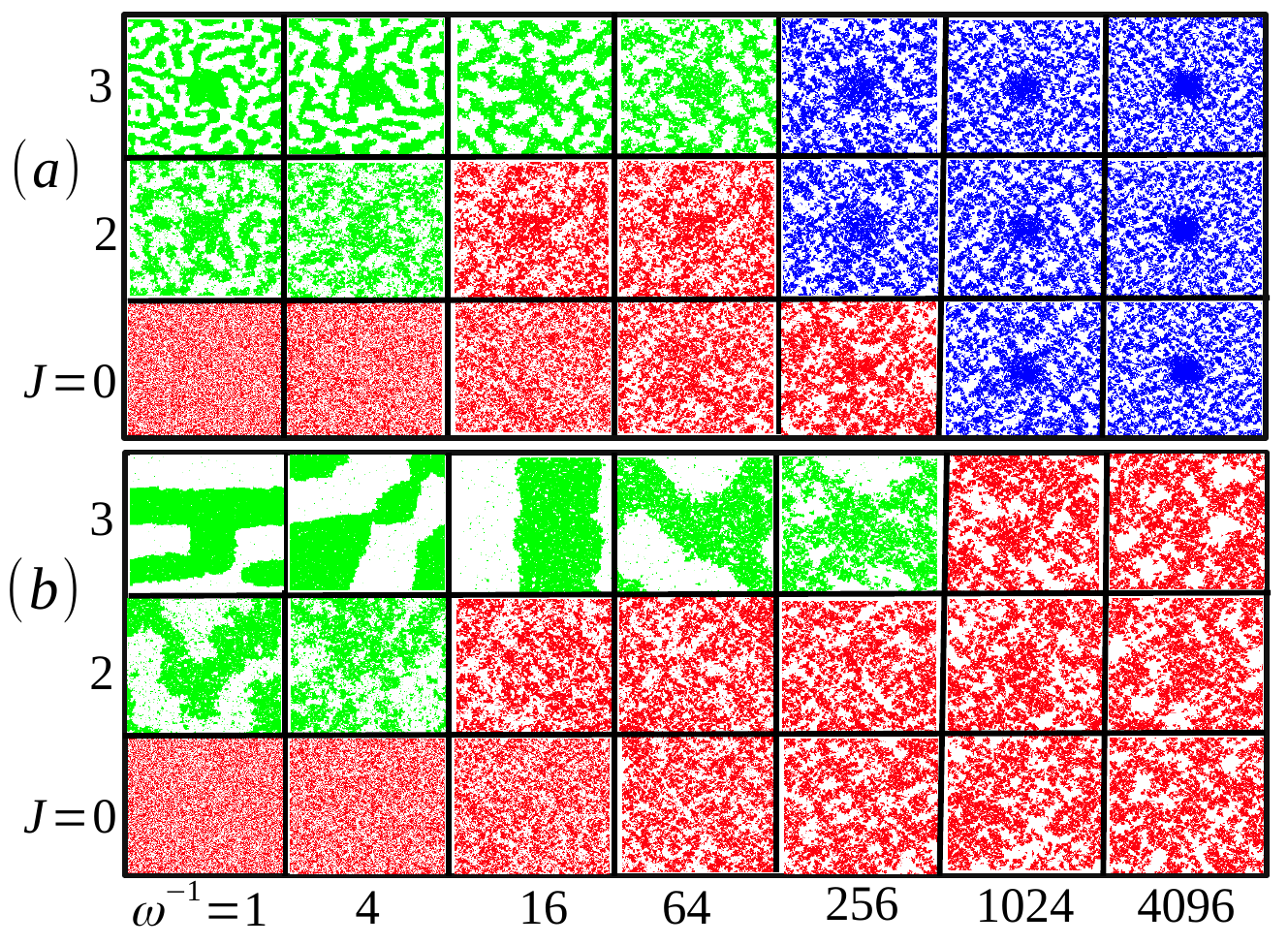}
\caption{ Snapshot of $2\times10^4$ RTPs on a $200\times 200$ square lattice at different $J$ and $\omega.$ Initial condition: $800$ particles form a square nucleation center, the rest are placed randomly. (a) At time $t=10^5$, MIPS-like phases appear for large $\omega^{-1}.$ (b) These transient states disappear at $t=10^7.$}
\label{fig:yesMIPS}
\end{figure}

In summary,  we find that  run and tumble particles  on a square lattice exhibit a phase separation transition   at $\rho=1/2,$ when the attractive inter-particle interaction crosses  a threshold  $J_c(\omega)$ that   increases  with increase of  tumble rate $\omega$ and  approach the  exactly known value of  lattice gas model  when $\omega \to \infty.$
This  transition belongs to Ising universality class.  A common feature   observed in  the  models we study here is that, in presence of attractive interaction,  the order parameter  $\bar \phi$ decreases  with  increase  of motility.  Such  behaviour  has been observed earlier  in numerical simulations of active particles  in continuum \cite{Cates2011}, but later studies  \cite{Redner, Reentrant_I, Reentrant_II, Reentrant_III} showed  that 
with sufficiently  high  motility a phase separated state re-appears   leading   to  a re-entrant  phase transition. The lattice models   we study here  do not exhibit  re-entrant phenomena.  To investigate this, we simulate the  IRTP model  on  a $200\times 200$ square lattice with $20,000$ particles.  Initially we  use $800$ particles  to form a  compact nucleation center  at the center of the lattice  and rest are distributed randomly  and uniformly.  The snapshots of the simulations for  different $\omega$     are shown in   Fig. \ref{fig:yesMIPS} for $J=0,2,3.$ At short times, as shown in Fig. \ref{fig:yesMIPS} (a)  where $t=10^5$ MCS, motility   initially hinders  cluster formation  when $J\ne0$ but  with increased motility  appears to cluster around  the  nucleation center, but they disappear   when relaxation time is increased to   $t= 10^7$  (Fig. \ref{fig:yesMIPS} (b)).

Surprisingly  for $J=0,$   as  shown  in  Fig. \ref{fig:yesMIPS} (b) and  in  Fig. \ref{fig:heat_map}  phase separation   is not  observed  even for  motility $\omega^{-1}$ as large as $4096.$  It is indeed a matter  to worry as  it is widely accepted in literature that in presence  of repulsion (here hard-core repulsion) persistently moving particles (both  active Brownian  particles  and   RTPs) exhibit motility induced  phase separation.   The reason why we don't observe MIPS  for $J=0$   could be   many-fold, which are listed below. \\

\noindent {\it (A) Recombination time-scale:}  For a phase-separated state to be stable, particles that break away from the cluster should join back in a short time. During the interval $\Delta t$, the probability that a single passive particle that has just left the cluster joins back is $p\propto \Delta t/4$ as it must choose to move in the direction of the cluster (out of four possibilities). For an RTP, however, it happens with a smaller probability $\propto \Delta t^2 \omega/3.$ This is because an RTP that has broken away from a cluster is oriented outward and thus, to join back, it must first reorient towards the cluster with probability $\omega \Delta t/3$ and subsequently move with probability $\propto \Delta t$. It is therefore unlikely (in fact, impossible in $\Delta t \to 0$ limit) that a departed RTP rejoins the same cluster in a short time. 

\begin{figure}[h]
\centering
\includegraphics[width=5.2cm]{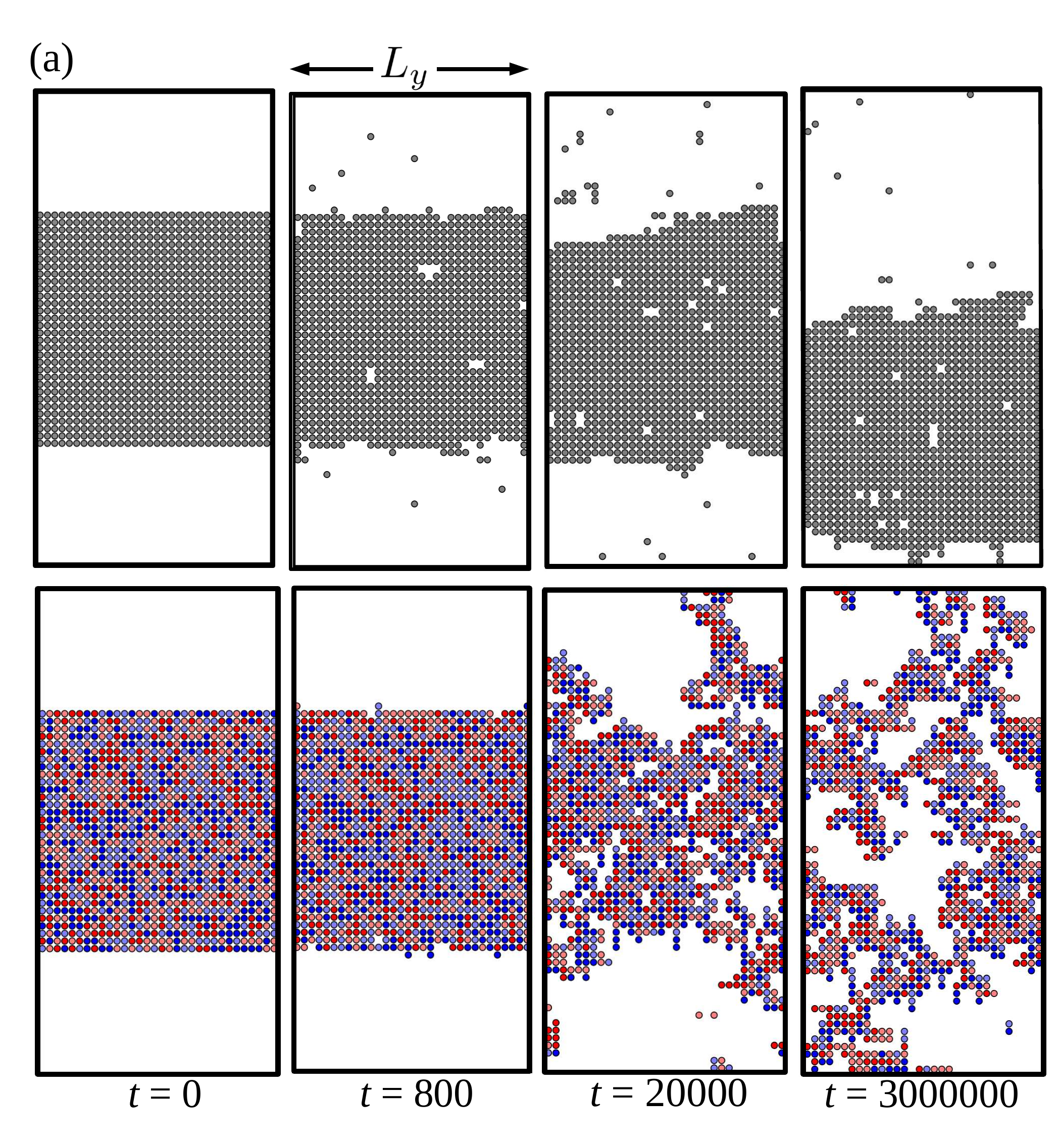} \hspace{0.3cm}
\includegraphics[height=5.2cm]{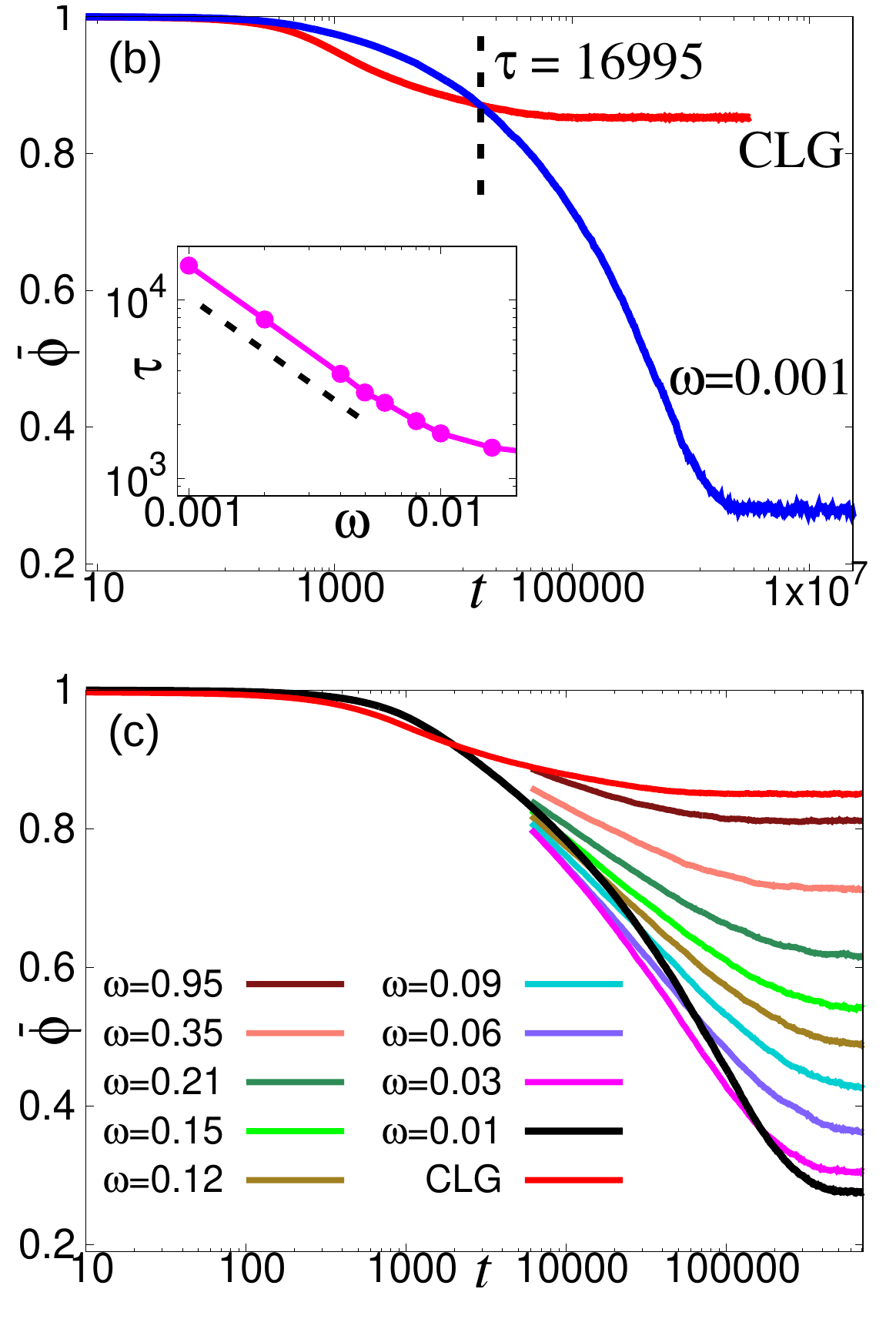}
\caption{(a) Evolution of a fully ordered configuration of usual CLG (upper panel) and IRTPs with $\omega=0.001$ (lower panel, colors represent different orientations) at $J=10/9.$ For IRTPs, colors represent different internal orientations.
In short times, $t<\tau,$ IRTPs maintain the order better than their non-motile counterparts. However, as $t\to\infty,$ CLG finds a steady state with very high order (as $T<T^{LG}_c=1.134$) compared to RTPs (where $J\gg J_c(\omega=0.001)\approx 7.14)$. (b) $\tau(\omega)$ is defined as the time where $\bar\phi$ of IRTPs with motility $\omega^{-1}$ becomes smaller than their non-motile counterpart, $\tau(0.001)=16995.$ Log scale plot of $\tau$ vs. $\omega$ along with a dashed line of slope $-1,$ in the inset, indicates that $\tau$ diverges as $\omega^{-1}.$ (c) $\bar\phi(t)$ for different $\omega$: systems with smaller $\omega$ take longer to reach the steady state.}
\label{fig:ising_HRTP_time}
\end{figure}

To investigate this in detail we start from a fully ordered configuration (as in Fig. \ref{fig:ising_HRTP_time}(a)) and observe how CLG evolves in comparison to RTPs having a small $\omega$. At short times, only particles residing on the coexistence line can leave the cluster, and the length of the 
coexistence line $\approx L$ does not change substantially. As a consequence,  particles in CLG leave the cluster at a constant rate. For RTPs, however, this rate decreases with time, because RTPs who could not leave the cluster at the first attempt, for not being oriented along the direction opposite to the coexistence line, are unlikely to tumble and reorient properly in the next attempt when $\omega$ is small. However, this scenario changes when $t$ crosses a characteristic time scale $\tau,$ when the departed particles in CLG rejoin the cluster by diffusion and keep the cluster in good health but departed RTPs keep on moving further away and disperse in the low-density zone because of their asynchronous noisy dynamics. Thus one expects $\bar \phi(t)$ for RTPs to be larger than that of CLG until $t <\tau,$ which is shown in Fig. \ref{fig:ising_HRTP_time}(b) for $\omega=0.001.$
 A typical evolution of CLG and IRTP with $\omega=0.001$ is compared in Fig. \ref{fig:ising_HRTP_time}(a) 
for increasing values of $t.$
Initially, at $t=800,$ the configuration of CLG appears more disordered than that of RTPs whereas the opposite happens as time progresses beyond $\tau;$ eventually RTPs reach a steady state with a much lower order, which is also observed from the steady state value of $\bar \phi$ in \ref{fig:ising_HRTP_time}(b) for $\omega=0.001$ (and in Fig. \ref{fig:ising_HRTP_time}(c) for other $\omega$ values.). Thus, for small $\omega$ one must evolve the RTP systems for longer times to reach the true steady state. Short-time simulation may lead to erroneous conclusions. Note that $\tau$ is only a comparative time scale; it must diverge as $\omega^{-1},$ the persistence time of RTP (verified in the inset of Fig. \ref{fig:ising_HRTP_time}(b)). The relaxation time (time required for the system to reach the steady state) is much larger than $\tau.$ \\

\noindent {\it (B) Absence of diffusion:}
Another crucial difference in comparison to earlier active matter dynamics is the absence of diffusivity.
On a lattice,  inter-particle distance cannot be made lower than the lattice  unit  and thus a neighbouring particle  pair   with internal orientations pointing towards  each other 
remain immovable until  one of them   tumble. Since  the 
average  time required  for a particle  to tumble is ${\cal O}( \omega^{-1}),$   eventually  RTPs   form  micro-clusters   
and remain in that jammed state for a long time  when  $\omega$ is small. Positional diffusion  can be added to the model  by allowing  particles  to move  with a smaller rate  in directions other than  their internal orientation; this diffusivity also rescues the system from getting stuck in ``micro-clusters".

Conventionally, the tuning parameter   of  MIPS transition  is  the  Peclet number which is the ratio  of  persistent length to diffusive length.  It is quite possible  that in absence  of positional diffusion the Peclet number  assumes an infinite value    leading to   absence of MIPS transition  in IRTP model at $J=0.$
\\

\noindent {\it (C) Insufficient repulsion:}  Another reason  for the absence of MIPS could be insufficient repulsion.  It is   generally believed that  finite-range  repulsive interaction  is  required for   MIPS  transition. In our model, the  attractive interaction  $J$ works between particles when they are separated by  one lattice unit whereas the   (infinite) repulsion   applies  when particles try to occupy the same site (range of repulsion is zero). Thus attraction  dominates for any $J>0$ and as a result, $J=0$  could be a  singular limit.

The  primary reason   responsible   for nonexistence of  phase separation transition   in  the IRTP model  at $J=0$ is not clear.
We are  investigating  all these  possible reasons in detail. 

Finally  we  end the discussion by asking a naive question, {\it what else is required to stabilize MIPS in these models?.}    Motility induced  phase separation  has been observed in other RTP models   in  absence of any attractive interaction. Dynamics of  these models  include additional features: either  the tumbling rate is not constant \cite{Kourbane-Houssene}, or the run rate (speed) decreases with increased local particle density $\rho_l$ \cite{cates_motility-induced_2015}, or particles can move in directions other than their internal orientation \cite{MIPS-Ising2D3, Whitelam}. The conclusions of this study cannot be extended directly to these special cases without scrutiny; the reasons follow. If the interaction radius of the system is $\eta,$ then $\rho_l$ must be defined by considering particles 
within a distance $R\gg \eta.$ It is then obvious that the movement of particles by one lattice unit does not change the 
 energy ($\Delta E=0$) when $\rho_l=0$ or $1;$ corresponding Metropolis rates or the speed $v(.)$ are the same. Thus, $v(\rho_l)$ is a non-monotonic function.  On the other hand, when RTPs are allowed to move in other directions besides their internal orientation \cite{MIPS-Ising2D3, Chate2020}, particles that depart from a well-formed cluster
may join back easily and stabilize it.  In models where tumbling rate $\omega$ depends on system size $L$ or coarsening  length scale $l,$  say $\omega \sim L^{-1},$ \cite{Kourbane-Houssene}, the persistence length of RTPs also grow as $\sim L.$  Thus particles in low-density regions may easily travel macroscopic distances to join the clusters in high-density zones and  produce MIPS transition in some cases.

\ack PKM would like to thank Urna Basu for the helpful discussions and careful reading of the manuscript. IM acknowledges the support of the Council of Scientific and Industrial Research, India in the form of a research fellowship (Grant No. 09/921(0335)/2019-EMR-I).

\end{document}